# Valley Splitting and Polarization by the Zeeman Effect in Monolayer MoSe$_2$

Yilei Li[1], Jonathan Ludwig[2], Tony Low[1], Alexey Chernikov[1], Xu Cui[3], Ghidewon Arefe[3], Young Duck Kim[3], Arend M. van der Zande[3], Albert Rigosi[1], Heather M. Hill[1], Suk Hyun Kim[1], James Hone[3], Zhiqiang Li[2], Dmitry Smirnov[2], Tony F. Heinz[1*]

[1] *Departments of Physics and Electrical Engineering, Columbia University, New York, NY 10027, USA*

[2] *National High Magnetic Field Laboratory, Tallahassee, FL 32312, USA*

[3] *Department of Mechanical Engineering, Columbia University, New York, NY 10027, USA*

\* To whom correspondence should be addressed: tony.heinz@columbia.edu

Abstract: We have measured circularly polarized photoluminescence in monolayer MoSe$_2$ under perpendicular magnetic fields up to 10 T. At low doping densities, the neutral and charged excitons shift linearly with field strength at a rate of $\mp$ 0.12 meV/T for emission arising, respectively, from the *K* and *K'* valleys. The opposite sign for emission from different valleys demonstrates lifting of the valley degeneracy. The magnitude of the Zeeman shift agrees with predicted magnetic moments for carriers in the conduction and valence bands. The relative intensity of neutral and charged exciton emission is modified by the magnetic field, reflecting the creation of field-induced valley polarization. At high doping levels, the Zeeman shift of the charged exciton increases to $\mp$ 0.18 meV/T. This enhancement is attributed to many-body effects on the binding energy of the charged excitons.

Monolayer MoSe$_2$ features two inequivalent valleys in the Brillouin zone of its electronic structure. The broken inversion symmetry of the monolayer allows this valley degree of freedom to be selectively accessed by optical helicity, providing a unique platform to probe and manipulate the charge carriers in the two valleys. [1-8] Since the valleys are linked by time-reversal symmetry, they are energetically degenerate, while the magnetic moments of the corresponding valley states are of the same magnitude, but have opposite sign [1,9,10]. Coupling to the valley magnetic moments by a magnetic field thus provides an attractive, but as yet unexplored method of breaking the valley degeneracy [11,12]. This presents new opportunities for the study of the fundamental physical properties of the valley electronic states, as well as for the development of new approaches to valleytronic control.

In this work, we experimentally investigate the ability of a perpendicular magnetic field to tune the valley energies in monolayer MoSe$_2$ by valley-resolved magneto-photoluminescence (magneto-PL) spectroscopy. *Lifting of the valley degeneracy* is demonstrated through the opposite energy shifts induced in the excitonic transitions in the two valleys by the magnetic field. The magnitude of this Zeeman shift, 0.12 meV/T, agrees with the predicted magnetic moments of the valley states. In the presence of a magnetic field, with split *K* and *K'* valleys, we create an *equilibrium valley polarization*, *i.e.*, an imbalance in the charge distribution in the two valleys, by doping the sample. This behavior is revealed by the variation of the relative emission intensity of the charged and neutral excitons. Further, by comparing the direction of the energy shift of the conduction band and the relative intensity of the negatively charged exciton, we are able to clarify the valley configuration of these bright trion states. In addition, the doping dependent trion Zeeman shift reveals the modification to the many-body binding energy by the creation of valley polarization.

MoSe$_2$ monolayers were prepared by mechanical exfoliation of bulk crystals on SiO$_2$ (300 nm)/Si substrates. Ti/Au contacts to the monolayer were fabricated using electron-beam lithography. With these electrical contacts, we could control the charge density of the sample by gating with respect to the Si substrate. The PL measurements were performed using excitation by a continuous-wave laser with a photon energy of 2.33 eV. The pump radiation was focused by an objective to a spot size of 3 μm on sample, with a power of ~100 μW. The reflected PL signal was collected by the same objective, and then filtered by a circular polarizer composed of a broadband quarter wave plate and a linear polarizer. The circularly polarized PL light was measured by a spectrometer equipped with a liquid-nitrogen cooled CCD detector. The PL measurements were carried out at a temperature of 10 K under a perpendicular magnetics field ranging from -10 T to +10 T.

In Figs. 1(a-b), we present the magnetic field dependence of the PL spectrum for the right ($\sigma_+$) and left ($\sigma_-$) circularly polarized photons in the low-doping regime. We see prominent and well-separated features from emission of neutral ($X^0$) and charged ($X^-$) excitons [13-30]. The $X^-$ peak lies ~30 meV below the $X^0$ peak, with the energy difference reflecting the trion binding energy. The emission energies of the neutral and charged excitons shift monotonically with magnetic field. The sign of the shift is reversed for photons of opposite circular polarization. Since the emitted photons with $\sigma_+$ and $\sigma_-$ polarization states correspond to optical transitions from the different valleys [2-4], the fact that they shift in opposite directions is clear evidence that the *K/K'* valley degeneracy has been lifted, *i.e.*, that valley splitting has been produced. The valley splitting is a direct consequence of the opposite magnetic moments associated with the bands in the two valleys, governed by the time reversal relation of the unperturbed valleys [5].

We extract the transition energy of the neutral exciton emission by fitting each spectral peak to a Lorentzian line shape. The resulting variation of the $X^0$ energy with the magnetic field is plotted in Figs. 1(c-d). Using a linear fit, we obtain a rate of the Zeeman shift for the neutral exciton of $-0.12$ meV/T and $+0.12$ meV/T for the $K$ and $K'$ valleys, respectively. The experimental uncertainty of the extracted Zeeman slope is $\pm 0.01$ meV/T based on multiple measurements of the same sample.

To analyze the measured Zeeman shifts, let us recall the relevant band structure of MoSe$_2$ near the $K/K'$ points. Both the conduction and valence bands, which have extrema at the $K/K'$ points, are split by spin-orbit interactions [31-33]. For MoSe$_2$ the highest valence band and the lowest conduction band are predicted to have the same spin and to give rise to the observed bright exciton state. The other spin-split bands, separated by ~30 meV and ~200 meV, respectively, from the conduction and valence bands are not accessed in our PL measurements, which are conducted at liquid-helium temperature and at modest levels of sample doping. Since the relevant conduction and valence bands are well described by a quadratic dispersion at the $K$ and $K'$ points [32], we analyze the magneto-optic response in a two-band effective mass framework throughout this paper. The significant excitonic effects are considered through the introduction of binding energies for these states. [13-30]

Within this framework, we now consider the influence of an applied magnetic field B. The energy of the neutral exciton $E_{X0}$ is given by the band gap, reduced by the exciton binding energy: $E_{X0} = E^c - E^v - E_{X0}^b$. Here $E^c$ and $E^v$ represent the energies of the conduction and valence band extrema; and $E_{X0}^b$ denotes the binding energy of the neutral exciton. Under an applied magnetic field, the exciton binding energies can acquire a diamagnetic shift, which is quadratic under in the B field [34,35]. Based on the observed linear variation of the transition energy with B, we infer, however, that this term is small. Hence, we conclude that the observed PL shift for the neutral exciton is simply due to the linear Zeeman response of the relevant conduction and valence band edges:

$$\Delta E_{X0} = \Delta E^c - \Delta E^v = -(\mu^c - \mu^v)B,$$

where $\mu^c$ and $\mu^v$ are the total magnetic moments of the conduction and valence bands.

We can then compare our experimental Zeeman shifts with the magnetic moments predicted on the basis of a two-band tight-binding description of the electronic states near the K and K' valleys [1]. In this model, the total magnetic moment of a charge carrier arises from contributions of the atomic orbital ($\mu_l$), carrier spin ($\mu_s$), and the delocalized Bloch wavefunction ($\mu_k$), which we denote as the wavefunction contribution. In the K valley, the contributions to the magnetic moments of the bands are $\mu_s^c = \mu_s^v = -\mu_B$, $\mu_l^c = 0$, $\mu_l^v = -2\mu_B$, $\mu_k^c = -\frac{m_0}{m_*^c}\mu_B$ and $\mu_k^v = -\frac{m_0}{m_*^v}\mu_B$, where $m_*^c$ and $m_*^v$ are the effective masses of the conduction and valence bands, and $\mu_B = 0.058$ meV/T is the Bohr magneton. The atomic orbital moments reflect the properties of the Mo 3d states with angular moment $2\hbar$ for the valence band and 0 for the conduction band that form the electronic states at the K point. The wavefunction contributions are equivalent to a Bohr magneton for each band, adjusted by the corresponding effective mass, and are of the same sign for both bands. All of the contributions in the K' valley are by time-reversal symmetry, exactly opposite in sign.

In Figs. 2(a-b), we illustrate the Zeeman shifts of the bands under the applied magnetic field, as well as the initial and final states for emission by a neutral exciton. Since interband optical transitions connect states in the conduction and valence bands with the same spin, the spin contribution to the Zeeman shifts cancel one another out in the transition energy. In the approximation of electron-hole symmetry, we have $m_*^c = m_*^v = m_*$ and $\mu_k^c = \mu_k^v = -\frac{m_0}{m_*}$. The wavefunction contributions to the Zeeman shifts then also cancel out in Zeeman shift of the transition energy [1]. For the *K* valley, we have accordingly a predicted Zeeman shift of the neutral exciton energy that depends only on magnetic moments of the Mo orbital angular momentum, namely, $\frac{\Delta E_{X0}}{\Delta B} = \mu^v - \mu^c = -2\mu_B = -0.116$ meV/T. The rather precise agreement between this model and our experiment is fortuitous: The slightly larger hole mass obtained within a three-band tight binding model would, for example, contribute ~0.02 meV/T to the predicted Zeeman shift [32]. The sign change of the Zeeman shift for the *K'* valley observed experimentally is, however, a robust theoretical prediction [10].

Now we turn our focus to the trion emission. In the low-doping regime, our experiment yields Zeeman shifts for the negative trion of $-0.12$ meV/T and $+0.12$ meV/T for the *K* and *K'* valleys, respectively [Figs. 1(c-d)]. Within experimental uncertainty, these values are the same as those found for neutral excitons. The trion emission energy can be expressed as $E_{X^-} = E^c - E^v - E_{X0}^b - E_{X^-}^b$, where $E_{X^-}^b$ denotes the binding energy of a trion, *i.e.*, the energy needed to dissociate it into a neutral exciton and a free electron [17,36]. The identical Zeeman slopes for $X^0$ and $X^-$ emission is a consequence of the weak influence of the magnetic field on $E_{X0}^b$ and $E_{X^-}^b$ at low doping levels.

Considering the B-field induced splitting of the *K* and *K'* valley energies together with the finite electron doping of the sample, we expect the creation of an imbalance in the charge distribution in the two valleys, *i.e.*, to the creation of static *valley polarization*. The presence of valley polarization is, in fact, clearly reflected in the relative intensities of the PL spectra of Fig. 1(a). Since we detect *valley-specific* emission through the polarization selection, the relative PL intensity of charged and neutral excitons, $I_{X^-}/I_{X0}$, provides information about the presence of charge carriers within a given valley (see Supplementary Material for a plot of $I_{X^-}/I_{X0}$). To understand this phenomenon in more detail, let us consider the possible valley configurations of the trion species. A negative trion species contributing to emission in the *K* valley consists of an electron and hole in the *K* valley, together with an additional electron that could be located either in the *K* or *K'* valley. Because of the correlation between electron spin and valley, these nominally degenerate states have different electron spin configurations [5], the intravalley trion corresponding to identical electron spins and the intervalley trion corresponding to opposite electron spins. Exchange interactions then lead to energy differences between these two valley configurations of the trion [31].

To determine whether one of the two possible valley configurations for the trion is favored over the other, let us examine trion emission from the *K* valley ($\sigma_+$). According to the magnetic moments in the two-band model [1], a positive magnetic field induces an upshift of the *K* valley conduction band with respect to the *K'* valley conduction band [Figure 2(c-d)]. As a result, in thermal equilibrium more electrons populate the conduction band in the *K'* valley than the *K* valley. This produces enhanced trion emission for the case of intervalley trions. For the case of intravalley trions, the situation is exactly reversed and the *K* valley trion emission will be reduced.

Experimentally, we see that $I_{X^-}/I_{X0}$ for the *K* valley increases with magnetic field [Fig. 1(a)]. We conclude that the trion emission originates predominantly from *intervalley* trions. This finding is compatible with recent theoretical prediction that the intervalley trion is ~6 meV more stable than the intravalley trion due to exchange interactions [31]. The intensity of the circularly-polarized trion emission thus not only provides a direct signature of the creation of valley polarization, but also provides important information about the valley character of the trion species itself.

We have also studied the Zeeman effect in the regime of high carrier doping. Through electrostatic gating, we increase the charge density in the sample by ~2.7×10$^{12}$ cm$^{-2}$, estimated from the gate capacitance and the applied voltage. Considering an electron band mass of 0.6 $m_0$ [32], this density of electrons would cause the Fermi energy $E^f$ to lie 11 meV above the conduction band minimum. Note that in this regime, the finite (10 K) sample temperature can be neglected. The corresponding dependence on magnetic field of the circularly-polarized PL spectra is shown in Figs. 3(a-b). In this high-doping regime, the neutral exciton peak is very weak, and we only analyze shifts in the trion emission energy. We again find a linear Zeeman shift for the trion, with opposite behaviors in the two valleys. However, the rate of the trion Zeeman shift at high doping is $\mp$ 0.18 meV/T for the *K* and *K'* valleys, respectively, compared with $\mp$ 0.12 meV/T for the same quantity in the regime of low charge density.

We now consider the mechanism leading to the observed 50% increase in the Zeeman shift of the trion in the high-doping regime compared to the low-doping regime. We attribute the changed behavior to the influence of the magnetic field on the trion binding energy $E^b_{X-}$. When the Fermi level is raised above the conduction band edge, $E^b_{X-}$ acquires a term due to a state-filling effect, in which additional energy is required to move the extra bound electron from the conduction band edge back to the Fermi level when the trion is dissociated, *i.e.*, $E^b_{X-}(E^f \geq E^c) = E^b_{X-}(E^f < E^c) + E^f - E^c$ [17,36].

Figs. 2(e-f) illustrate the Zeeman shifts of the bands, as well as the initial and final states of the trion emission process in the high-doping regime. The conduction band splitting results in charge transfer from one valley to the other and the production of valley polarization, as described above. Based on the total magnetic moment of the conduction band in the two-band model, we deduce an electron transfer between the two valleys of $\Delta n/\Delta B \approx 3.9\times10^{10}\,\text{cm}^{-2}/\text{T}$, assuming an electron effective mass of $0.6\,m_0$ [32]. Despite the valley charge transfer process, the Fermi energy, we note, remains unchanged since the density of states in the two valleys is equal.

The Zeeman shift of the trion emission at high doping can be analyzed quantitatively based on the band alignments shown in the Figs. 2(e-f) and the variation of the binding energy for the (intervalley) trions with doping in the *K'* valley described above. We find that the conduction-band contribution to the trion Zeeman shift cancels out, leaving only the valence-band contribution:

$$\Delta E_{X^-} = -\Delta E^v = \mu^v B, \qquad\qquad\qquad (E^f > E^c)$$

Within the two-band model, we then expect a Zeeman slope for the trion emission in the *K* valley of $\frac{\Delta E_{X^-}}{\Delta B} = \mu^v = \mu_l^v + \mu_s^v + \mu_k^v = -\left(2 + 1 + \frac{m_0}{m^*}\right)\mu_B$. Using $m^* = 0.6\,m_0$ [32] as before, we obtain a Zeeman slope of $-0.27$ meV/T. This model captures both the sign of the trion shift and the larger magnitude in the trion shift compared to the low-doping case, although it does not provide quantitative agreement with the measured value of -0.18 meV/T. The difference between the experiment and model presumably reflects the influence of the field-induced intervalley charge transfer on many-body corrections to the trion transition energy beyond the trion binding energy considered above.

In summary, through measurements of circularly polarized photoluminescence from neutral and charged excitons in MoSe$_2$, we have demonstrated how a perpendicular magnetic field lifts the *K/K'* valley degeneracy and also creates a valley polarization in doped samples. We have identified the intervalley configuration as the lower-energy state for the trion and have discussed the importance of many-body corrections in the observed variation in the trion emission energy with doping. The controlled lifting of the valley degeneracy by a magnetic field and corresponding creation of a steady-state valley polarization open up exciting opportunities to probe and utilize the valley degree of freedom in monolayer TMDCs. By tuning the valley transition energies, it should, for example, be possible to create optically excited states with intervalley quantum beats, as well as steady-state valley selective currents.

The authors acknowledge the U.S. Department of Energy, Office of Basic Energy Sciences for support through Columbia Energy Frontier Research Center (grant DE-SC0001085) for the preparation and characterization of samples and device structures; the National Science Foundation through grants DMR-1122594 and DMR-1124894 for support for the optical measurements and through grant DMR-1106225 for data analysis. H. M. H. and A. R. were supported by the NSF through an IGERT Fellowship (DGE-1069240) and through a Graduate Research Fellowship, respectively. A. C. acknowledges funding from the Alexander von Humboldt Foundation through a Feodor-Lynen Fellowship. J.L. acknowledges the support by NHMFL UCGP No. 5087. A portion of this work was performed at the National High Magnetic Field Laboratory, which is supported by National Science Foundation Cooperative Agreement No. DMR-1157490, the State of Florida, and the U.S. Department of Energy. The authors would like to thank Igor Aleiner, Timothy C. Berkelbach, and Junichi Okamoto for fruitful discussions.

**Figures:**

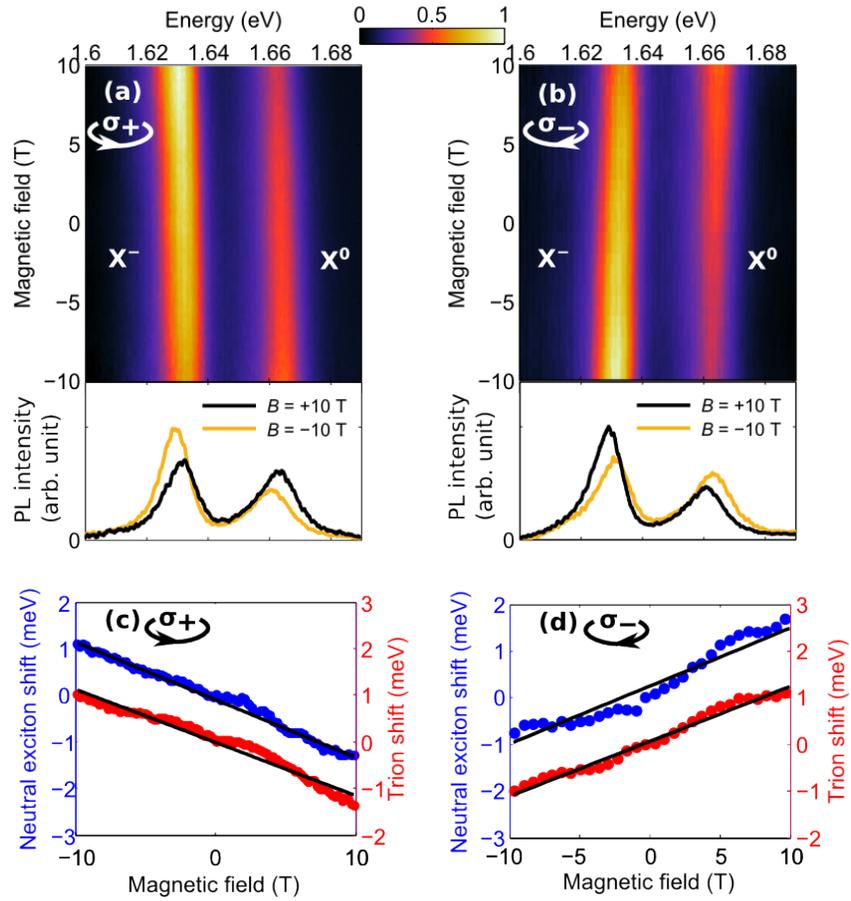

Figure 1. (a) and (b) False color representation of the $\sigma_+$ and $\sigma_-$ polarized PL spectra of monolayer MoSe$_2$ in the low-doping regime as a function of the strength of the applied perpendicular magnetic field. The spectra for fields of ±10 T are presented below the color map. The spectra are normalized to yield the same integrated intensity for the neutral and charged excitons, with a smooth background subtracted. (c) and (d) Extracted Zeeman shift of the neutral and trion energies for $\sigma_+$ and $\sigma_-$ PL. The solid lines are linear fits to the experimental data.

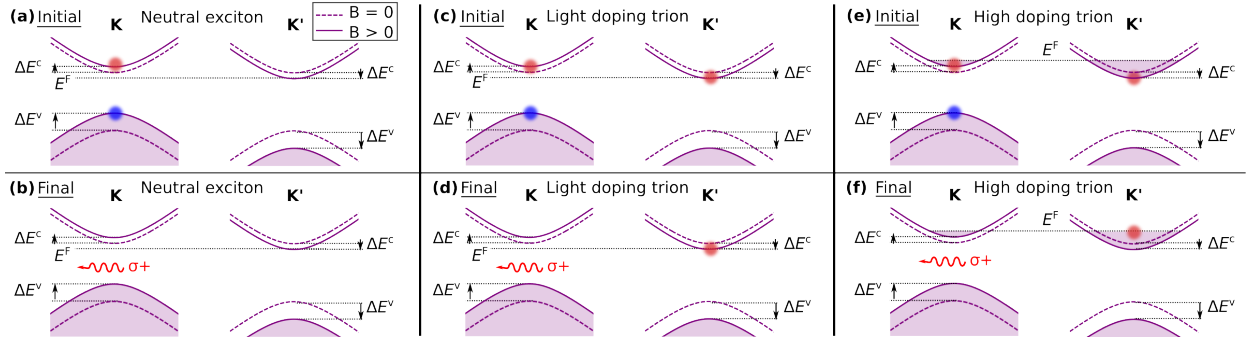

Figure 2. Schematic of valley splitting and polarization from the Zeeman effect and the initial and final states of the neutral and charged exciton emission process. In the presence of an out-of-plane magnetic field, the conduction and valence bands shift according to their respective magnetic moments. The dashed lines denote the bands at $B = 0$ and the solid lines correspond to the bands for $B > 0$. (a) and (b) The initial and final states of the neutral exciton emission process in the low-doping regime. (c) and (d) The initial and final states of the charged exciton emission process in the low-doping regime. (e) and (f) The initial and final states of the trion emission process in the high-doping regime.

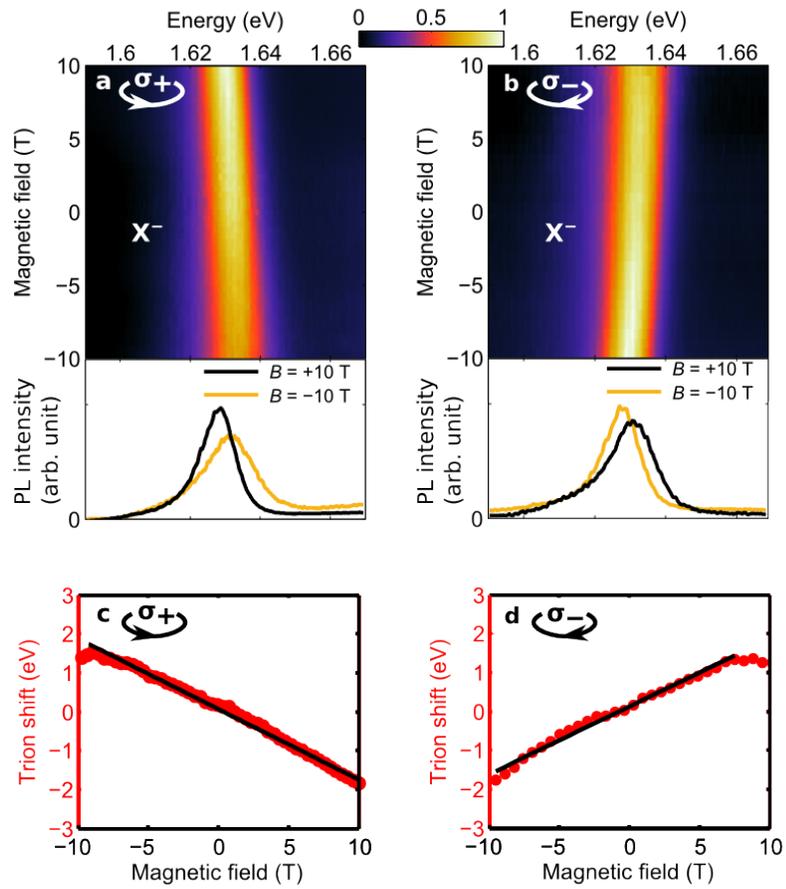

Figure 3. Photoluminescence of monolayer MoSe$_2$ in the high-doping regime as a function of applied magnetic field. All information is as in Fig. 1, except only the negative trion feature is observable and no data is available for the neutral exciton.